\documentclass[twocolumn,showpacs,amsmath,amssymb]{revtex4}
\usepackage{graphicx}
\usepackage{dcolumn}
\usepackage{bm}

\newcommand{\ba}{\begin{eqnarray}}
\newcommand{\ea}{\end{eqnarray}}
\newcommand{\ban}{\begin{eqnarray*}}
\newcommand{\ean}{\end{eqnarray*}}
\newcommand{\be}{\begin{equation}}
\newcommand{\ee}{\end{equation}}
\newcommand{\bd}{\begin{displaymath}}
\newcommand{\ed}{\end{displaymath}}
\newcommand{\ds}{\displaystyle }

\newcommand{\n}[1]{\label{#1}}
\newcommand{\nn}{\nonumber}

\newcommand{\lap}{\bigtriangleup}
\newcommand{\Rot}{\mbox{curl}\, }
\newcommand{\Div}{\mbox{div}}
\newcommand{\pa}{\partial}

\newcommand{\hh}{\, ,\hspace{0.5cm}}
\newcommand{\hhh}{\, ,\hspace{0.2cm}}

\newcommand{\BM}[1]{{\mbox{\boldmath $#1$}}}

\begin{document}



\title{Gravitational field of relativistic gyratons}

\author{Valeri P. Frolov${}^{(a)}$}
\email{frolov@phys.ualberta.ca}
\author{Werner Israel${}^{(b)}$}
\email{israel@uvic.ca}
\author{Andrei Zelnikov${}^{(a,c)}$}
\email{zelnikov@phys.ualberta.ca}
\address{
$^{(a)}$Theoretical Physics Institute,
  Department of Physics, University of Alberta,
  Edmonton, AB, Canada, T6G 2J1\\
$^{(b)}$Department of Physics and Astronomy, University of Victoria,
Canada, V8W 3P6\\
$^{(c)}$ Lebedev Physics Institute,
  Leninsky prospect 53, 119991, Moscow Russia
}

\date{\today}

\begin{abstract}   
The metric ansatz (\ref{2.1}) is used to describe
the gravitational field of a beam-pulse of spinning radiation
(gyraton) in an arbitrary number of spacetime dimensions $D$. First
we demonstrate that this metric belongs to the class of metrics for
which all scalar invariants constructed from the curvature and its
covariant derivatives vanish. Next, it is shown that the vacuum
Einstein equations reduce to two linear problems in
$(D-2)-$dimensional Euclidean space. The first is to find the static
magnetic potential ${\bf A}$ created by a point-like source. The
second  requires finding the electric potential $\Phi$ created by a
point-like source surrounded by given distribution of the electric
charge. To obtain a generic gyraton-type solution of the vacuum
Einstein equations it is sufficient to allow the coefficients in the
corresponding harmonic decompositions of solutions of the linear
problems to depend arbitrarily on  retarded time $u$ and substitute
the obtained expressions in the metric ansatz. These solutions are
generalizations of the gyraton metrics found in \cite{FrFu:05}. We
discuss  properties of the solutions for relativistic gyratons and
consider special examples. 
\end{abstract}

\pacs{04.70.Bw, 04.50.+h, 04.20.Jb \hfill Alberta-Thy-08-05}

\maketitle

\section{Introduction}
\n{s1}

Studies of the gravitational fields of beams and pulses of light have
a long history. Tolman \cite{To} found a solution in the linearized
approximation. Peres \cite{Pe1,Pe2} and Bonnor \cite{Bo} obtained
exact solutions of the Einstein equations for a pencil of light.
These solutions belong to the class of  pp-waves.   The
generalization of these solutions to the case where the beam of
radiation carries angular momentum   has been found recently in
\cite{FrFu:05}. Such a solution corresponds to a pulsed beam of
radiation with negligible radius of cross-section, finite duration in
time, and which has finite both energy $E$, and angular momentum $J$.
An ultra-relativistic source with these properties is called a
gyraton. 

The gyraton-type solutions are of general interest, since
for example they allow one to address the question: What is the
gravitational field of a photon or ultrarelativistic electron or
proton? This question becomes important in the discussion of possible
mini-black-hole production in future collider or cosmic ray
experiments. In the absence of spin, one can use the  Aichelburg-Sexl
metric \cite{AiSe,BaHo} to describe the gravitational field of each
of the colliding particles. Such an approach allows one to estimate
the cross-section for mini-black-hole formation
\cite{EaGi:02,YoNa:02,YoNa:03,YoRy} (for a general review see
\cite{Kant}). The metric obtained in \cite{FrFu:05} makes it possible
to consider the gravitational scattering and mini-black-hole
formation in the interaction of particles with spin. The estimates
show  \cite{FrFu:05} that the spin-spin and spin-orbit interaction
may be important at the threshold energies for mini-black-hole
formation. 

In the present paper we study the gravitational field of gyratons. We
start by discussing the general properties of the metric (\ref{2.1})
describing the gravitational field of relativistic gyratons
(Section~\ref{s2}).  First we show that  this metric belongs to the
class of metrics for which all the scalar invariants constructed from
the curvature and its covariant derivatives vanish identically. For
${\bf A}=0$ this result was obtained by Amati and Klimcik
\cite{AmatiKlimcik:89} and Horowitz and Steif
\cite{HorowitzSteif:90},  who argued that such metrics are classical
solutions to string theory. (For a general discussion of spacetimes
with vanishing curvature invariants see \cite{Coley,PPCM,NSI}).

After this we show that the vacuum Einstein equations for  metric
(\ref{2.1})  in a spacetime with arbitrary number $D$ of dimensions
reduce to the linear problems for the gravitoelectric ($\Phi$) and
gravitomagnetic (${\bf A}$) potentials in the $(D-2)-$dimensional
Euclidean space. These linearized problems can be easily solved. The
solutions obtained in  \cite{FrFu:05} are characterized by the
property that only lowest harmonics are present in the harmonic
decomposition of $\Phi$  and ${\bf A}$. For this reason one can
consider the gyraton solutions presented in \cite{FrFu:05} as some
ground state, while the more general solutions obtained in this paper
are their excitations (or distortions). It should be emphasized that
the vacuum solutions are valid only outside the region occupied by
gyratons. In order to obtain a solution describing the total
spacetime one needs to obtain a solution inside the gyraton. This
solution depends on the gyraton structure. In the present paper we do
not discuss concrete gyratons models. But since we obtain a general
solution for the vacuum metric outside a gyraton, one can gurantee
that for any model of the gyraton there exists a corresponding
solution, so that the characteristics of the gyraton are "encoded" in
the parameters of the exterior vacuum metric.

After discussing the asymptotic properties of the gyraton metrics
(Section~\ref{s3}), we consider general solutions for $4-$
(Section~\ref{s4}) and $5-$dimensional (Section~\ref{s5})  gyraton
metrics. Section~\ref{s6} discusses the higher dimensional gyraton
metrics. In Section~\ref{s7} we summarize the obtained results and
discuss open problems.

\section{Metric for relativistic gyratons}
\n{s2}

\subsection{Gyraton metric ansatz}

Let us consider Brinkmann \cite{Br} metric in $D-$dimensional
spacetime of the form
\[
ds^2=g_{\mu\nu}dx^{\mu} dx^{\nu}
\]
\be\n{2.1}
=-2~du~dv+d{\bf x}^2+\Phi~du^2
+2~({\bf A},d{\bf x})du\, ,
\ee 
\be
\Phi=\Phi(u,{\bf x})\hh
A_a=A_a(u,{\bf x})\, .
\ee
Evidently, $l^{\mu}\pa_{\mu}=\pa_v$ is the null Killing vector.

When $\Phi={\bf A}=0$, the coordinates $x^1=v=(t+\xi)/\sqrt{2}$ and
$x^2=u=(t-\xi)/\sqrt{2}$ are null. The coordinate $u$ remains null
for the metric (\ref{2.1}). The metric
is generated by an object moving with the velocity of light in the
$\xi-$direction. The coordinates   $(x^3,\ldots,x^D)$ are 
coordinates of an $n-$dimensional space ($n=D-2$) transverse to the
direction of motion. We use bold-face symbols to denote vectors in
this space. For example, ${\bf x}$ is a vector with components $x^a$
($a=3,\ldots,D$). We denote by $r$ the length of this vector,
$r=|{\bf x}|$. We also denote
\be
d{\bf x}^2=\sum_{a=3}^{D} (dx_a)^2\hh
({\bf A},d{\bf x})=\sum_{a=3}^{D}A_a~dx^a\, ,
\ee
\be
\lap =\sum_{a=3}^{D} \pa_a^2\hh
\Div{\bf A}=\sum_{a=3}^{D}A^a_{,a}\, .
\ee
Later we assume that the sum is taken over the repeated indices and
omit the summation symbol. Working in the Cartesian coordinates we
shall not distinguish between upper and lower indices.

The form of the metric (\ref{2.1}) is invariant under the following
(gauge) transformation
\be\n{gauge}
v\to v+\lambda(u,{\bf x})\hhh
A_a\to A_a-\lambda_{,a}\hhh
\Phi\to \Phi-2\lambda_{,u}\, .
\ee
It is also invariant under rescaling
\be
u\to au\hhh
v\to a^{-1}v\hhh
\Phi\to a^2 \Phi\hhh
{\bf A}\to a {\bf A}\, .
\ee

It is easy to show that for the metric (\ref{2.1})
\be\n{g}
\sqrt{-g}=1\, ,
\ee
and the inverse metric is
\be
g^{\mu\nu}\pa_{\mu}\pa_{\nu}=-(\Phi -{\bf A}^2)\pa_v^2-2\pa_u\pa_v
+2A_a\pa_a\pa_v+\pa_{a}^2\, .
\ee

The non-vanishing components of the Christoffel symbol
$\Gamma_{\mu,\nu\lambda}$ are
\[
\Gamma_{v,\mu\nu}=\Gamma_{\mu,\nu v}~=~\Gamma_{a,bc}~=0\hh
\Gamma_{u,uu}={1\over 2}\partial_u\Phi\, ,
\]
\be\n{Gl}
\Gamma_{a,uu}=\partial_uA_a-{1\over 2}\partial_a\Phi \hh
\Gamma_{u,ua}={1\over 2}\partial_a\Phi \, ,
\ee
\[
\Gamma_{u,ab}={1\over 2}(\partial_aA_b+\partial_bA_a)\hh
\Gamma_{a,bu}=-{1\over 2}F_{ab}\, ,
\]
where
\be
F_{ab}=\pa_a A_b-\pa_b A_a\, .
\ee
We shall also need the Christoffel symbols
$\Gamma^{\mu}_{\nu\lambda}=g^{\mu\alpha}\Gamma_{\alpha,\nu\lambda}$.
Their non-vanishing components are
\[
\Gamma_{uu}^v=-{1\over 2}(\Phi_{,u}+A^a\Phi_{,a})+A^a A_{a,u}\, ,
\]
\be\n{G}
\Gamma_{ua}^v=-{1\over 2}(\Phi_{,a}-F_{ab}A^b)\hhh
\Gamma_{ab}^v=-{1\over 2}(A_{a,b}+A_{b,a})\, ,
\ee
\[
\Gamma_{uu}^a=A^a_{\ ,u}-{1\over 2} \Phi^{,a}\hhh
\Gamma^b_{ua}={1\over 2} F_a^{\ \ b}\, .
\]

It is easy to check that 
\be\n{pp}
l_{\mu;\nu}=0\, .
\ee
It means that the null Killing vector ${\bf l}$ is covariantly
constant. In the 4-dimensional case, space-times admitting a
(covariantly) constant null vector field are called plane-fronted
gravitational waves with parallel rays, or briefly pp-waves (see e.g.
\cite{EK,J,ES}). Similar terminology is often used for higher
dimensional metrics (see e.g. \cite{B,O}).

\subsection{Curvature invariants}

In the next section we derive conditions under which metric
(\ref{2.1}) is Ricci flat and hence obeys the vacuum Einstein
equations. But before this let us prove that the metric (\ref{2.1}) 
belongs to the class of metrics with vanishing curvature invariants. 
Namely, all the local scalar invariants constructed from the Riemann
tensor and its covariant derivatives for the metric (\ref{2.1})
vanish. This statement is valid {\it off shell}, that is the metric
need not be a solution of the vacuum Einstein equations. This
property is well known for 4-dimensional case, since pp-wave
solutions are of Petrov type N. Generalization of this result to
higher-dimensional metrics (\ref{2.1})  with ${\bf A}=0$ was given in
\cite{AmatiKlimcik:89,HorowitzSteif:90}. (For a general discussion of
spacetimes with vanishing curvature invariants see
\cite{Coley,PPCM,NSI}).

To demonstrate that curvature invariants vanish for the metric
(\ref{2.1}), let us consider a covariant tensor $A_{\mu \ldots \nu}$.
We shall call such a tensor {\em degenerate} if it has the following
properties: It does not depend on $v$, and its components, which
either contain at least one index $v$ or do not contain index $u$,
vanish. Since $\pa_v$ is the Killing vector, the Riemann curvature
tensor does not depend on $v$. Using the expressions for the
Christoffel symbols (\ref{G}) one can show that  the only
non-vanishing components of the Riemann tensor are $R_{[au][bu]}$,
$R_{[ab][cu]}$, and  $R_{[cu][ab]}$. Hence it is degenerate. Let us
demonstrate now that the action of a covariant derivative
$\nabla_{\mu}$ on a degenerate tensor $A_{\mu \ldots \nu}$ gives a
tensor which is also degenerate. Really, since
$\Gamma^{\alpha}_{v\mu}=0$ one has
\be
\nabla_v A_{\mu \ldots \nu}=\pa_v A_{\mu\ldots
\nu}-\Gamma^{\alpha}_{v\mu}A_{\alpha\ldots \nu} -\ldots
\Gamma^{\alpha}_{v\nu}A_{\mu \ldots \alpha}=0\, .
\ee
Thus the covariant differentiation of the degenerate tensor cannot
have a non-vanishing $v$ component. Since $\Gamma^u_{\mu\nu}=0$, the
covariant differentiation cannot also produce a non-vanishing component
which does not contain index $u$. 

Consider now a scalar invariant constructed from any set of
degenerate tensors and $g^{\mu\nu}$. The only non-vanishing component
of $g^{\mu\nu}$ which contains an index $u$ is $g^{uv}=-1$. Hence a
scalar invariant constructed  from degenerate tensors and metric
always vanishes.

\subsection{Calculation of the Ricci tensor}

In order to calculate the Ricci tensor for the metric (\ref{2.1}) let
us introduce the following vectors
\be
V^{\lambda}={\pa x^{\lambda}\over \pa v}\hh
U^{\lambda}={\pa x^{\lambda}\over \pa u}\hh
e_{(a)}^{\lambda}={\pa x^{\lambda}\over \pa x^a}\, .
\ee
One has
\be
V^{\beta}_{\ ;\alpha}=\Gamma^{\beta}_{v\alpha}=0\hh
U^{\beta}_{\ ;\alpha}=\Gamma^{\beta}_{u\alpha}\hh
e_{(a);\beta}^{\lambda}=\Gamma^{\lambda}_{a\beta}\, ,
\ee
\be\n{ue}
U^{\beta}_{\ ;\alpha}e_{(c);\beta}^{\alpha}=
\Gamma^{\beta}_{u\alpha}\Gamma^{\alpha}_{c\beta}=
\Gamma^{b}_{ua}\Gamma^{a}_{cb}+\Gamma^{u}_{ua}\Gamma^{a}_{cu}=0\, .
\ee
The last equality holds because
$\Gamma^{a}_{cb}=\Gamma^{u}_{ua}=0$.

The Ricci identity implies
\be\n{ricci}
R_{\alpha\beta}Y^{\alpha} X^{\beta}=
(X^{\beta}_{\ ;\alpha}Y^{\alpha})_{;\beta}-
X^{\beta}_{\ ;\alpha}Y^{\alpha}_{\beta}-X^{\beta}_{\ ;\beta\alpha}Y^{\alpha}\,
.
\ee

From the relation $V^{\beta}_{\ ;\alpha}=0$ it follows that
$R_{\alpha\beta}Y^{\alpha} V^{\beta}=0$ and hence
\be
R_{v\alpha}=0\, .
\ee
Let us set $Y^{\alpha}=e^{\alpha}_{(a)}$ and $X^{\beta}=U^{\beta}$,
then using (\ref{ue}) one has
\be
R_{au}=(U^{\beta}_{\ ;\alpha}e^{\alpha}_{(a)})_{;\beta}-U^{\beta}_{\
;\beta\alpha}
e^{\alpha}_{(a)}\, .
\ee
Using (\ref{g}) one obtains
\be\n{ubb}
U^{\beta}_{\ ;\beta}={1\over \sqrt{-g}}
\pa_{\beta}(\sqrt{-g}\delta^{\beta}_u)=0\, .
\ee
One also has
\be
U^{\beta}_{\ ;\alpha}
e^{\alpha}_{(a)}=\Gamma^{\beta}_{ua}={1\over 2}\delta^{\beta}_{b}
F_a^{\ \ b}  
-{1\over 2}\delta_v^{\beta}(\Phi_{,a}-F_{ab}A^b)
\, .
\ee
Since $\Phi$ and $A_a$ do not depend on $v$, and
$\Gamma^{\alpha}_{v\beta}=0$, one has
\be
R_{au}={1\over 2} \pa_b F_a^{\ \ b}\, .
\ee

Similarly
\be
R_{uu}=(U^{\beta}_{\ ;\alpha} U^{\alpha})_{;\beta}
-U^{\beta}_{\ ;\alpha}U^{\alpha}_{\ ;\beta}
-U^{\beta}_{\ ;\beta\alpha}U^{\alpha}\, .
\ee
Relation (\ref{ubb}) implies that the last term on the right hand side
vanishes. Since  $U^{\beta}_{\ ;\alpha}
U^{\alpha}=\Gamma^{\beta}_{uu}$ using (\ref{G}) one obtains
\be
(U^{\beta}_{\ ;\alpha} U^{\alpha})_{;\beta}=
\pa_b (A^b_{\ ,u}-\Phi^{,b})\, .
\ee
One also has
\be
U^{\beta}_{\ ;\alpha}U^{\alpha}_{\ ;\beta}=
\Gamma^{\beta}_{u\alpha} \Gamma^{\alpha}_{u\beta}=
\Gamma^{b}_{ua} \Gamma^{a}_{ub}=-{1\over 4}F_{ab}F^{ab}\, .
\ee
Combining these results one obtains
\be
R_{uu}=\pa_u \Div {\bf A}-{1\over 2}\lap \Phi +{1\over 4}{\bf F}^2\, ,
\ee
where
\be
\Div {\bf A}=\pa_a A^a\hh
{\bf F}^2=F_{ab}F^{ab}\hh
\lap \Phi =\pa_a \pa^a \Phi\, .
\ee

Finally, let us substitute $X^{\alpha}=e^{\alpha}_{(a)}$ and 
$Y^{\beta}=e^{\beta}_{(b)}$ into (\ref{ricci}), then one has
\be
R_{ab}=\left( e^{\beta}_{(b);\alpha} e^{\alpha}_{(a)}\right)_{;\beta}-
e^{\beta}_{(b);\alpha} e^{\alpha}_{(a);\beta}-
e^{\beta}_{(b);\beta\alpha}e^{\alpha}_{(a)}\, . 
\ee
Using (\ref{G}) it is easy to show that
\be
e^{\beta}_{(b);\alpha} e^{\alpha}_{(a)}=\Gamma^{\beta}_{ab}=-{1\over
2} \delta_v^{\beta} (A_{a,b}+A_{b,a})\, ,
\ee
\be
e^{\beta}_{(b);\alpha} e^{\alpha}_{(a);\beta}=\Gamma^{\beta}_{b\alpha}
\Gamma^{\alpha}_{a\beta}=0\hh
e^{\beta}_{(b);\beta}=\pa_{\beta}\delta^{\beta}_b=0\, .
\ee
Hence $R_{ab}=0$.

\subsection{Vacuum equations for gravitational field of a gyraton}

To summarize, the metric (\ref{2.1}) is a solution of vacuum Einstein
equations if and only if the following  equations are satisfied
\be\n{e1}
\pa_b F_a^{\ \ b}=0\, ,
\ee
\be\n{e2}
\lap \Phi -2\pa_u \Div {\bf A}={1\over 2}{\bf F}^2\, .
\ee
In the next section it will be shown that for solutions describing a
gyraton with finite energy and angular momentum the quantities 
$\Phi_{,a}$ and $F_{ab}$ are vanishing at transverse space infinity. 
We assume that the
homogeneous equations are valid everywhere outside the point ${\bf
x}=0$ where a point-like source is located. It is easy to see that
the left hand side of (\ref{e2}) is gauge invariant, that is
invariant under the transformations (\ref{gauge}). In the `Lorentz'
gauge $A^a_{,a}=0$  equation (\ref{e2}) takes the form
\be\n{e2a}
\lap \Phi={1\over 2}{\bf F}^2\, .
\ee
Here ${\bf F}^2=F_{ab}F^{ab}$. Using the analogy of gravity with
electromagnetism, one can say that the problem of solving the
$D-$dimensional vacuum Einstein equations for the gyraton metric is
reduced to finding an electric potential $\varphi$ and magnetic field
$F_{ab}$ created by a local source in the $(D-2)-$dimensional
Euclidean space. For these solutions the retarded time $u$ plays a
role of an external parameter which enters through the dependence of
point-like sources on $u$.

As we already mentioned in the Introduction, in physical applications
there always exists a source of the gravitational field which
generates the metric (\ref{2.1}). We called this source a gyraton
\cite{FrFu:05}. In order to obtain a solution describing the total system,
one must obtain a solution inside the region occupied by the gyraton
and to glue it together with a vacuum solution (\ref{2.1}) outside
it. In the present paper we study only solutions outside the
gyraton.  We shall obtain a {\em general} solution of the
magnetostaic equations (\ref{e1}) for point-like currents localized
at the point ${\bf x}=0$. Since this equation is linear this current
can be written as a linear combination of  $\delta({\bf x})$ and its
derivatives. Similarly, one can write a general solution of the
equation 
\be\n{el}
\lap \varphi=0\, ,
\ee
with a charge density, localized at ${\bf x}=0$, or, what is
equivalent, with the charge-density proportional to  $\delta({\bf
x})$ and its derivatives. It is convenient to write $\Phi=\varphi
+\psi$ , where
\be\n{eli}
\lap \psi={1\over 2}{\bf F}^2\, .
\ee

After finding $A_a(u,{\bf x})$ and $\varphi(u,{\bf x})$, one needs
only to find the `induced' potential $\psi(u,{\bf x})$ determined by
the equation (\ref{eli}). A formal solution of this problem can be
obtained as follows. The Green function of the $n-$dimensional
Laplace operator
\be\n{gf}
\Delta {\cal G}_n({\bf x},{{\bf x}'})
=-\delta ({\bf x}-{{\bf x}'})
\ee
is
\be\n{gf2}
{\cal G}_2({\bf x},{{\bf x}'})=-{1\over 2\pi}\ln |{\bf x}-{\bf x}'|\,
, \ \ \mbox{if  }n=2\, ,
\ee
\be
{\cal G}_n({\bf x},{{\bf x}'})={g_n\over |{\bf x}-{{\bf x}'}|^{n-2}}\, , 
\ \ \mbox{if  }n>2\,
,
\ee
\be
g_n={\Gamma( { n-2\over 2})\over 4\pi^{n/2}} \, .
\ee
Using these Green functions  one can present the solution for
$\psi$ in the form
\be\n{psi}
\psi(u,{\bf x})=-{1\over 2} \int d{\bf x}' {\cal G}_n({\bf x},{{\bf
x}'}) {\bf F}^2(u,{\bf x}')\, .
\ee

Let us emphasized that in the general case the solution (\ref{psi}) is
only formal and may not have a well-defined sense. The reason is that
for a point-like current, ${\bf F}$ has a singularity at ${\bf x}=0$.
If one considers this singular function as a distribution, one needs
to define what is the meaning of ${\bf F}^2$ in (\ref{psi}). This
problem does not exist for a distributed source (gyraton). If we do
not want to input an explicit form of the matter distribution within
gyraton, we can proceed as follows \cite{remark}. 

Suppose ${\bf F}$ and $\varphi$ are solutions with localized sources.
Let us surround a point ${\bf x}=0$ by a $(D-3)-$dimensional surface
$\sigma$. For example, one may choose $\sigma$ to be a round
$(D-3)-$dimensional sphere of small radius $\epsilon$. Denote by
$F^{(\sigma )}_{ab}=F_{ab} \vartheta(\sigma)$, where
$\vartheta(\sigma)$ is equal to 1 outside $\sigma$ and vanishes
inside $\sigma$. The "magnetic" field $F^{(\sigma )}_{ab}$ obeys the
equation
\be\n{ee1}
\pa_b F_a^{(\sigma )\ \ b}=-n_b F_{a}^{\ \ b} \delta(\sigma)\, ,
\ee
where ${\bf n}$ is the unit normal to $\sigma$ vector directed to the
exterior of  $\sigma$. In orther words, the field  $F^{(\sigma
)}_{ab}$ corresponds to the special case of an extended gyraton for
which its angular momentum density is localised on $\sigma$. The value
$\psi^{(\sigma)}$ obtained for  $F^{(\sigma )}_{ab}$ by using
(\ref{psi}) is well difined. Certainly this function $\psi^{(\sigma)}$
depends on the choice of $\sigma$. Suppose $\sigma'$ is another
surface, surrounding ${\bf x}=0$, and lying inside $\sigma$. It is
easy to see that outside $\sigma$ one has
\be
\lap (\psi^{(\sigma)}-\psi^{(\sigma)})=0\, .
\ee
That is outside $\sigma$ these two solutions
$\psi^{(\sigma)}$ and $\psi^{(\sigma)}$ differ by a term which can be
absorbed into the solution $\varphi$. 

For a distributed source (a gyraton) one can use a similar procedure.
If one is interested in the gravitational field of the gyraton
outside a surface $\sigma$ surrounding the matter distribution one
can calculate $\psi^{(\sigma)}$ and choose $\varphi$ correspondingly.
For a given distribution of the gyraton matter, the parameters of the
vacuum solution outside the gyraton are uniquely specified. In
sections \ref{s4} and \ref{s5} we shall give explicit examples of the
vacuum gyraton solutions.

\section{Energy and angular momentum of a gyraton}
\n{s3}

\subsection{Weak field approximation}

The asymptotics of functions $\Phi$ and $A_a$ at the
transverse-spatial infinity are related to the energy and angular
momentum of a gyraton. In order to find these relations let us
consider a linearized problem. Let us write the Minkowski metric in
the form
\be
ds_0^2=\eta_{\mu\nu}dx^{\mu} dx^{\nu}=-2du\, dv +d{\bf x}^2 \, .
\ee
Then its perturbation $h_{\mu\nu}$ generated by the stress-energy tensor
$T_{\mu\nu}$ obeys the equation
\be\n{pert}
\Box h_{\mu\nu}=-\kappa \bar{T}_{\mu\nu}\hh
\bar{T}_{\mu\nu}=
(T_{\mu\nu} -{1\over n}\eta_{\mu\nu} T)\, ,
\ee
where $\kappa=16\pi G$. 

In the general case, if the metric (\ref{2.1}) is a solution of the
Einstein equations, then the  stress-energy tensor which generates
this solution possesses the following properties: (1) Its non-vanishing
components are $T_{uu}$ and $T_{ua}$; (2) These components do not
depend on $v$; and (3) It obeys the conservation law $T_{\mu\nu}^{\ \
;\nu}=0$. The latter condition in the linear approximations reduces to
the relation
\be\n{lcon}
T_{ua,a}=0\, . 
\ee
The metric perturbation $h_{\mu\nu}$ generated by such a
stress-energy tensor also does not depend on $v$. Thus instead of
$D$-dimensional $\Box$ operator in (\ref{pert}) one can substitute
the $n$-dimensional flat Laplace operator $\lap$
\be\n{pert_a}
\lap h_{\mu\nu}=-\kappa {T}_{\mu\nu} \, .
\ee
We omit the {\it bar} over ${T}_{\mu\nu}$ since its trace vanishes.

Using the Green function (\ref{gf}) one can write the following
expression for ${h}_{\mu\nu}({\bf x})$
\be
{h}_{\mu\nu}(u,{\bf x})=\kappa \int d{{\bf x}'}
{\cal G}_n(|{\bf x}_{\perp}-{{\bf x}'}|)\bar{T}_{\mu\nu}(u,{{\bf x}'})\,
.
\ee

\subsection{Metric asymptotics}

Consider ${T}_{\mu\nu}(u,{\bf x})$ as a function of ${\bf x}$ and
suppose that it vanishes outside of some compact region. We denote by $l$
the size (in the transverse direction) of this region. To determine
the field at far distance $r\gg l$ we use the following relation
\be
|{\bf x}-{{\bf x}'}|\sim r-{ ({\bf x},{{\bf x}'})\over r}\, .
\ee
Thus if one point, ${\bf x}$, is at far distance from the source, while
the other is close to it one has the following asymptotics for the
Green functions
\be
{\cal G}_2({\bf x},{{\bf x}'})=-{1\over 2\pi}\ln r +{({\bf x},{\bf
x}')\over 2\pi r^2}+\ldots\,
, \ \ \mbox{if  }n=2\, ,
\ee
\be
{\cal G}_n({\bf x},{{\bf x}'})={g_n\over r^{n-2}}
+{g_n(n-2)({\bf x},{\bf x}')\over r^n}+\ldots
\, , \ \ \mbox{if  }n>2\,
,
\ee
where $\ldots$ denote the terms of higher order in $1/r$.
Similarly, one has
\be
h_{\mu\nu}=-{\kappa\over 2\pi}\ln r {\cal T}_{\mu\nu}+
{\kappa\over 2\pi r^2}  x^a {\cal J}_{a\mu\nu}+\ldots\,
, \ \ \mbox{if  }n=2\, ,
\ee
\be
h_{\mu\nu}={\kappa g_n\over r^{n-2}}{\cal T}_{\mu\nu}
+{\kappa g_n (n-2)\over r^{n}} x^a {\cal J}_{a\mu\nu}+\ldots\,
, \ \ \mbox{if  }n>2\, ,
\ee
where
\be
{\cal T}_{\mu\nu}=\int d{\bf x} T_{\mu\nu}\hh
{\cal J}_{a\mu\nu}=\int d{\bf x} x_a\, T_{\mu\nu}\, .
\ee
The structure of the stress-energy tensor implies that only
components $h_{uu}$ and $h_{ua}$ do not vanish. In order to
relate the coefficients, which enter the asymptotic expressions for
these components, to physical quantities such as energy and angular
momentum we use the following relations
\be\n{pr1}
\int d{\bf x} T_{ua}=-\int d{\bf x} T_{uc}^{\ \ ,c}x_a=0\, ,
\ee
\be\n{pr2}
\int d{\bf x} (T_u^a x^b+T_u^b x^a)=-\int d{\bf x} T_{u\ ,c}^c x^a
x^b=0\, .
\ee
We used here the conservation law (\ref{lcon}).

The energy $E$ and the angular momentum $J_{ab}$ in the flat spacetime
are
defined by the relation
\be
E=\int d\xi d{\bf x} T_{tt}\hhh
J_{ab}=\int d\xi d{\bf x} (T_{ta}x_b-T_{tb}x_a)\, .
\ee
The integration is performed over a surface $t=$const. At this surface
$d\xi=-\sqrt{2} du$, thus
\be
\int_{-\infty}^{\infty} d\xi (\ldots)=\sqrt{2}\int_{-\infty}^{\infty}
du (\ldots)\, .
\ee
One also has
\be
T_{tt}={1\over 2}T_{uu}\hh
T_{ta}={1\over \sqrt{2}}T_{ua}\, .
\ee
By combining these relations one obtains
\be
E=\int du \varepsilon(u)\hh
J_{ab}=\int du j_{ab}(u)\, ,
\ee
\be
\varepsilon(u)={1\over \sqrt{2}}\int d{\bf x} T_{uu}\hhh
j_{ab}(u)=2\int d{\bf x} T_{ua} x_b\, .
\ee
Relation (\ref{pr2}) shows that $j_{ab}=-j_{ba}$. The function
$\varepsilon(u)$ describes the energy-density profile of the gyraton
as a function of the retarded time $u$, while $j_{ab}(u)$ is  similar
profile functions for the components of the density of the angular
momentum.

Using these results one obtains
\be\n{huu}
\Phi \sim h_{uu}= \kappa \sqrt{2} \varepsilon\left\{ 
\begin{array}{c} -{\ds 1\over \ds 2\pi}\ln r\, ,\quad\mbox{if  }n=2\, ,\\
\\
\displaystyle {g_n\over r^{n-2}}\, ,\quad\mbox{if  }n>2\, ,
\end{array}
\right.
\ee
\be\n{hua}
A_a\sim h_{ua}= {\kappa  g_n(n-2) j_{ab}x^b\over r^n}\, .
\ee
The latter relation is valid in the 4-dimensional spacetime (for
$n=2$) if one substitutes $1/(2\pi)$ for  $g_n (n-2)$.

\subsection{Canonical form}

If $j_{ab}$ were a time independent antisymmetric matrix then by
making rotations 
\be\n{tran}
{x}^a=O^a_{\ b} \tilde{x}^b\hh
{x}_a= \tilde{x}_c O^c_{\ a}
\ee
one would be able to bring $h_{ua}$ into a form where instead of
$j_{ab}$ stands its block canonical form \cite{Gant}
\be\n{can}
\tilde{j}_{ab}= \left( \begin{array}{ccccc}
0 &j_1& 0& 0&\ldots\\
-j_1 & 0& 0& 0&\ldots\\
0 & 0& 0& j_2&\ldots\\
0 & 0&-j_2& 0&\ldots\\
\ldots&\ldots&\ldots&\ldots&\ldots
\end{array} \right)\, .
\ee

In the presence of time dependence the situation is slightly more
complicated. Let us consider transformations (\ref{tran}) with time
dependent orthogonal matrix $O^a_{\ b}(u)$. Then
\be
d{x}^a=O^a_{\ b} d\tilde{x}^b+\dot{O}^a_{\ b} \tilde{x}^b du\hhh
d{x}_a=O^c_{\ a} d\tilde{x}_c+\dot{O}^c_{\ a} \tilde{x}_c du\, ,
\ee
where $\dot{B}=\pa_u B$.
Under these transformations the metric (\ref{2.1}) preserves its form
with
\be
\tilde{A}_a=A_b O^b_{\ a}+B_{ab} \tilde{x}^b\, ,
\ee
\be
\tilde{\Phi}=A_a \dot{O}^a_{\ b}\tilde{x}^b
+C_{ab}\tilde{x}^a\tilde{x}^b\, ,
\ee
where
\be
B_{ab}=O_{ac}\dot{O}^c_{\ b}=-\dot{O}_{ac} {O}^c_{\ b}\, ,
\ee
\be
C_{ab}=\dot{O}_{ac}\dot{O}^c_{\ b}=-B_{ac}B^{c}_{\ b}\, .
\ee
It is easy to see that $B_{ab} \tilde{x}^b$ is itself a solution of
the magnetostatic equations and correspond to a constant magnetic
field with $F_{ab}=-B_{ab}$. It means that the (linearly growing at
infinity) terms generated by time dependent  rigid rotations can be
compensated by adding to $A_a$ a new solution corresponding to a
constant magnetic field. This is a direct analogue of the Larmor
theorem in gravitomagnetism \cite{Mash:93}.  

To summarize, we demonstrated that by making a time dependent rotation
and adding to $A_a$ a vector potential for a homogeneous time
dependent magnetic field it is always possible to transform a solution
(\ref{2.1}) into the form where $\Phi$ and $A_a$ have the 
asymptotics (\ref{huu}) and (\ref{hua}),
where $j_{ab}$ is an  antisymmetric matrix in its canonical block form
(\ref{can}).

\section{$4-$dimensional gyratons}
\n{s4}

\subsection{General solution}

Before analyzing general gyraton-like solutions in an arbitrary
number of spacetime dimensions, we consider special
lower-dimensional cases.

Let us derive a gyraton metric in a 4-dimensional spacetime. In this case
the number of transverse dimensions $n=2$ and our problem reduces to
2-D electro- and magnetostatics.

Let us consider the equation (\ref{e1}) for the magnetic field. Any
antisymmetric tensor of the second order in a 2-dimensional space
can be written as $F_{ab}=Fe_{ab}$, where $e_{ab}$ is the totally
antisymmetric tensor. Substituting this representation into (\ref{e1})
one obtains that $F=$const. It is easy to see that the corresponding
vector potential $A_a$ can be written as
\be
A_3=\alpha x^4\hh A_4=\beta x^3\hh
F=\beta-\alpha\, .
\ee
The gauge transformation (\ref{gauge}) with $\lambda=\gamma x^3 x^4$
changes the coefficients $\alpha\to \alpha -\gamma$ and $\beta\to
\beta -\gamma$ but preserves the value $F$.

Equation (\ref{eli}) takes the form
\be
\lap \psi={1\over 2} F^2\, .
\ee
If $F\ne 0$, the solution $\psi$ does not vanish at infinity. We
exclude this case. Thus we put $F=0$. 

Let us choose a 2-dimensional contour surrounding the source at ${\bf
x}=0$. When $F_{ab}=0$, the value of the integral 
\be\n{circ}
j(u)={2\over \kappa}\oint_{C} A_a dx^a\hh
j(u)={1\over 2}\epsilon^{ab}j_{ab}\, ,
\ee
does not depend on the choice of the contour $C$. This quantity which
enters the solution (\ref{2.1}) has the meaning of angular momentum
of the gyraton. In polar coordinates
$(r,\phi)$
\be
x^3+ix^4=r e^{i\phi}\, .
\ee
the corresponding potential $A_a$ can be written as
\be\n{abp}
A_r=0\hh
A_{\phi}={\kappa\over 4\pi}j(u)\, .
\ee

Let us consider now equation (\ref{el}) for the 2-dimensional
`electric' potential $\varphi$. A solution corresponding to a
point-like charge is
\be
\varphi_0 = -{\kappa \sqrt{2}\over 2\pi} \varepsilon(u) \ln r\, .
\ee
Any other solution of this equation
decreasing at infinity can be written as
\be\n{ad}
\varphi=\varphi_0+\sum_{n=-\infty}^{'\infty} {b_n \over r^{|n|}}
e^{in\phi}\hh
\bar{b}_{n}=b_{-n}\, .
\ee
$\sum'$ indicates that the term $n=0$ is excluded.
In the electromagnetic analogy, the harmonics with $n\ge 1$ describe
the field created by an electric $n$-pole. Since $F=0$,
$\Phi=\varphi$ and the solution for a distorted gyraton in
$4-$dimensional spacetime is
\begin{eqnarray}\nn
ds^2&=&-2~du~dv+dr^2+r^2 d\phi^2 +{\kappa\over 2\pi} j(u) du d\phi 
\\
\n{4d}
&+&\left[ -{\kappa \sqrt{2}\over 2\pi} \varepsilon(u) \ln r
+\varphi\right] du^2 \, ,
\end{eqnarray}
where $\varphi=\varphi(u,r,\phi)$ is given by (\ref{ad}) with $b_n=b_n(u)$.

It should be emphasized, that the solution (\ref{4d}), which is a
special case of (\ref{2.1}), is the pp-wave metric. The properties of
pp-wave metrics in 4-dimensional spacetime are well known (see e.g.
\cite{J,EK,ES}). In particular, a vacuum pp-wave metric in a simply
connected region can be written in the form where ${\bf A}=0$. In a
case of a gyraton, because of the presence of a sigularity at ${\bf
x}=0$, the gauge transformations (\ref{gauge}) cannot be used to
banish the potential ${\bf A}$ {\em globally}. The situation here is
similar to the well known Aharonov-Bohm effect \cite{AB,Ham}. The
topological invariant $j(u)$, which has the meaning of the density of
the angular momentum of the gyraton, is similar to the magnetic flux
in the Aharonov-Bohm case. 

In study of the Aharonov-Bohm effect it is usually  helpful to consider
at first a tube of finite radius where the magnetic field is
localized. Similarly, in order to obtain well defined and finite
expression for the gyraton metric, one must consider a spreaded
source of  finite size. In the 4-dimensional case the procedure
proposed in section \ref{s2} does not work. We describe now a simple
model of an extended gyraton in the 4-dimensional case. Let us modify
the expression for (\ref{abp}) as follows
\be
A_r=0\, ,\ 
A_{\phi}={\kappa\over 4\pi}j(u) \left[{r^2\over
r_0^2}\vartheta(r_0-r)+ \vartheta(r-r_0)  \right]\, .
\ee
For this modified vector potential the field strength $F_{ab}=e_{ab}
F$ is
\be
F={\kappa\over 2\pi}{j(u)\over r_0^2}\vartheta(r_0-r)\, .
\ee 
In other words, the field strength $F$ is constant inside a disc of
radius $r_0$. Outside this disc the field vanishes, while the
contour integral (\ref{circ}) is $j(u)$ as earlier. 

The function $\psi$ for such an extended gyraton can be found by using
(\ref{gf2}) and (\ref{psi}). In polar coordinates one has
\be
\psi={F^2\over 4\pi}\int_0^{r_0} dr' r' Q\, ,
\ee
\be
Q=\int_0^{2\pi} d\phi \ln (r^2+{r'}^2-2rr'\cos\phi)\, .
\ee
For $r>r_0$ one has $Q=4\pi \ln r$. Thus
\be
\psi= {1\over 2}F^2 r_0^2 \ln r \, . 
\ee
It means that outside the gyraton $\psi$ has  the same form as
$\varphi_0$ and can be absorbed into the latter by renormalizing the
function $\varepsilon(u)$. 

\subsection{Boosted 4-$D$ NUT metric}

As an aside, it is worth mentioning that the Aichelburg-Sexl boost
\cite{AiSe} of the NUT stationary vacuum geometry \cite{NUT} is a
particular member of the class (\ref{4d}).

A convenient symmetric form of the NUT metric is
\[
ds^2={dr^2\over f(r)}+(r^2+a^2)\, (d^2\theta+\sin^2\theta\, d\phi^2)
\]
\be\n{nut}
-f(r)(dt-2a\cos\theta d\phi)^2\, ,
\ee
where 
\be
f(r)={r^2-2mr-a^2\over r^2+a^2}\, .
\ee
It is known \cite{Bonnor,MaRu:05} that the (singular) source of this
geometry consists of a pair of semi-infinite line sources along the
axis ($\theta=0$ and $\theta=\pi$ respectively), endowed with equal
and opposite average angular momenta $\pm a/2$ per unit length, and
joined to a massive particle at the origin. The two line sources are
massless to linear order in $a$, i.e., to within terms of the order
of the gravitational potential energy, which cannot be localized
unambiguously.

To boost the metric (\ref{nut}), it is sufficient to consider its
linearized form
\[
ds^2=(1+{2m\over r})(d\rho^2+\rho^2 d\phi^2+d{\bar
z}^2)
\]
\be
+4a{\bar{z}\over r} d\phi d{\bar t}- (1-{2m\over r}) d\bar{t}^2\, ,
\ee
where $r^2=\rho^2+\bar{z}^2$. We apply a Lorentz transformation
\be
\bar{z}={1\over 2}(v\, e^{-\chi}-u e^{\chi})\hh
\bar{t}={1\over 2}(v\, e^{-\chi}+u e^{\chi})\, ,
\ee
where $u=t-z$ and $v=t+z$. In the limit $\chi\to\infty$ this sends the
originally static source moving along the path $z=t$ in the new
frame. 
Noting that $\lim_{\chi\to\infty} {\bar{z}/r}=\epsilon(u)=\pm 1$,
rescaling mass and angular momentum according to
\be
m=\mu e^{\chi}\hh a=\alpha e^{\chi} \, ,
\ee
and using the distributional identity
\be\n{lim}
\lim_{\chi\to\infty}{e^{\chi}\over \sqrt{\rho^2+u^2
e^{2\chi}}}={1\over |u|}-\delta(u)\ln |\rho/l|\, ,
\ee
where $l$ is an arbitrary length scale, we readily obtain the limiting
form
\be\n{bnut}
ds^2=d\rho^2+\rho^2 d\phi^2-du dv -2\mu \delta(u)\ln |\rho/l|du^2
-2\alpha \epsilon(u) du d\phi\, .
\ee
(In (\ref{bnut}), we have absorbed the term $1/|u|$ from (\ref{lim})
by a transformation of $v$.)This is a special case of (\ref{4d}). It
represents a pair of semi-infinite gyratons with equal and opposite
angular momentum densities $\alpha \epsilon(u)$, joined to the
Aichelburg-Sexl boosted particle of energy $\mu$. Classification of
the 4-dimensional pp-waves with an impulsive profile based on their
symmetries can be found in \cite{AiBa}.

\section{$5-$dimensional gyratons}
\n{s5}

\subsection{General solution}

In order to obtain a solution for the gravitational field of a
5-dimensional gyraton one needs to analyze electro- and magnetostatics
in a flat 3-dimensional space. 

Let us consider first the `magnetic' equation (\ref{e1}). Using the
standard 3-dimensional notations one can write these equations in the
form
\be\n{m3}
{\bf B}=\Rot {\bf A}\hh
\Rot  {\bf B}=0\, .
\ee
The second equation implies that there exists a function $\Upsilon$, the
magnetic scalar potential, such that the magnetic field $B$ is
\be
{\bf B}=-\nabla \Upsilon\, .
\ee 
The first of the equations (\ref{m3}) implies that the magnetic
potential obeys the following equation
\be\n{up}
\lap \Upsilon =0\, .
\ee
Let $(r,\theta, \phi)$ be the spherical coordinates
\be
x^3+i x^4=r \sin\theta e^{i\phi}\hhh
x^5=r\cos\theta \, .
\ee
Then the general solution of (\ref{up}) decreasing at infinity can be
written as follows
\be
\Upsilon=\sum_{l=0}^{\infty} \sum_{m=-l}^{l}
a_{lm} {Y_{lm}(\theta,\phi)\over r^{l+1}}\, ,
\ee
where the complex coefficients $a_{lm}$ obey the conditions
$\bar{a}_{lm}=a_{l\, -m}$. Here $Y_{lm}(\theta,\phi)$ are spherical
harmonics
\be
Y_{lm}(\theta,\phi)=\sqrt{{ (2l+1)(l-m)!\over 4\pi (l+m)!}}
P_l^m(\cos\theta) e^{im\phi}\, .
\ee

The magnetic induction vector ${\bf B}$ is
\be
{\bf B}=-\sum_{l=0}^{\infty} \sum_{m=-l}^{l}
a_{lm} \nabla \left({Y_{lm}\over r^{l+1}}\right)\, .
\ee

In order to find the corresponding vector-potential ${\bf A}$ one
needs to solve the following equation
\be
\Rot {\bf A}=-\nabla \Upsilon\, .
\ee
It can be done by using the properties of vector spherical harmonics. 
Let us denote
\be
\BM{\Psi}_{lm}(\theta,\phi)=r\nabla Y_{lm}(\theta,\phi)\, ,
\ee
\be
\BM{\Phi}_{lm}(\theta,\phi)={\bf r}\times \nabla Y_{lm}(\theta,\phi)\,
.
\ee
The vector spherical harmonics obey the following relations \cite{BaEsGi}
\be\n{ra}
\nabla \times \left({\BM{\Phi}_{lm}\over r^{l+1}}\right)=\nabla \times
\left( { {\bf r}\times \BM{\Psi}_{lm}\over r^{l+2}}\right)\, .
\ee
\be\n{da}
\nabla \cdot \left({\BM{\Phi}_{lm}\over r^{l+1}}\right)=0\, .
\ee
Using the first of these relations one finds
\be\n{a}
{\bf A}={\bf A}_0-\sum_{l=1}^{\infty} \sum_{m=-l}^{l}
{a_{lm}\over l} {\BM{\Phi}_{lm} \over
r^{l+1}}\, .
\ee
The relation (\ref{da}) shows that the solution (\ref{a}) obeys the
following gauge condition
\be
\Div {\bf A}=0\, .
\ee
We denote by ${\bf A}_0$ a vector potential for $l=0$ case which
requires a special treatment, since in this case $\BM{\Phi}_{0}=0$ and
ratio ${\BM{\Phi}_{lm}/ l}$ is not determined.

A general solution of the equation (\ref{el}) for $\varphi$ can be
written as
\be
\varphi =\sum_{l=0}^{\infty} \sum_{m=-l}^{l}
b_{lm} {Y_{lm}(\theta,\phi)\over r^{l+1}}\, ,
\ee
where the coefficients $b_{lm}$ obey the conditions
$\bar{b}_{lm}=b_{l\, -m}$. For a gyraton solution
coefficients $a_{lm}$ and $b_{lm}$ are arbitrary functions of the
retarded time $u$. To obtain $\psi$ one can use (\ref{psi}) with the
\be
{\cal G}({\bf x},{\bf x}')={1\over 4\pi |{\bf x} -{\bf x}'|}\, .
\ee

\subsection{Monopole solution}

As we mentioned, the case of a magnetic monopole ($l=0$) is special.
Let us consider it in more details. The magnetic potential $\Upsilon$
for the magnetic monopole is
\be
\Upsilon=-{\mu\over r}\, ,
\ee
where $\mu$ is an arbitrary function of $u$. 
The magnetic induction vector has components
\be
B_r={\mu\over r^2}\hh
B_{\theta}=B_{\phi}=0\, .
\ee
The corresponding vector potential is of the form
\be
A_r=A_{\theta}=0\hh
A_{\phi}=-\mu \cos \theta\, .
\ee
The potential obeys the condition $\Div {\bf A}=0$ and the potential
$\psi$ is
\be
\psi ={\mu^2\over 4r^2}\, .
\ee
The corresponding monopole solution for the gyraton is
\[
ds^2=-2~du~dv+dr^2+r^2 d\theta^2 +r^2\sin^2\theta d\phi^2
\]
\be\n{5mon}
+(\varphi+{\mu^2(u)\over 4r^2})du^2 -2~\mu(u)\cos\theta du d\phi\, .
\ee
Here $\varphi=\varphi(u,r,\theta,\phi)$ is a solution of the equation
(\ref{el}).

Similarly to the 4-D case this metric is related to the boosted
NUT-like metric. Consider a metric ($R^2=r^2+w^2$)
\[
ds^2=-\left(1-{2m\over R^2}\right)dt^2+4a\cos\theta d\phi dt+
\]
\be
\left(1+{2m\over R^2}\right) (dr^2+r^2 d\theta^2 +r^2\sin^2\theta
d\phi^2+dw^2)\, .
\ee
It is Ricci-flat to linear order and it is a linearized version of 5-D
NUT spacetime. (An exact counterpart of this linearized metric does
not seem to be known.) Applying the boost
\be
w={1\over 2}(u e^{\chi}-v e^{-\chi})\hh
t={1\over 2}(u e^{\chi}+v e^{-\chi})\, 
\ee 
rescaling $m e^{2\chi}={1\over 2}\mu^2$, $ae^{\chi}=-\mu$, and noting
\be
\lim_{\chi\to\infty} R^{-2}=r^{-2}\, ,
\ee
we recover (\ref{5mon}) with $\varphi=1$ in the limit $\chi\to\infty$.

\subsection{Dipole solution}

The spherical harmonics for $l=1$ case are
\be
Y_{10}=\alpha {x^5\over r}\hh
\alpha=\sqrt{3\over 4\pi}\ee
\be
Y_{11}=\bar{Y}_{1-1}=-{\alpha\over \sqrt{2}} {x^3+ix^4\over r}\,.
\ee
Notice that ${\bf r}\times \nabla F(r)=0$. Using this property we
obtain
\be
\BM{\Phi}_{10}={\alpha\over r}(x^4,-x^3,0)\, ,
\ee
\be
\BM{\Phi}_{11}=-i{\alpha\over \sqrt{2}r}(x^5,-ix^5,x^3+ix^4)\, .
\ee
Let us denote
\be
\omega^{ab}=x^a dx^b -x^b dx^a\, .
\ee
Then the expressions for $({\bf A}_{1m}, d{\bf x})$ take the form
\be
({\bf A}_{10}, d{\bf x})=a_{10}{\alpha\over r^3}\omega^{34}\, ,
\ee
\be
({\bf A}_{11}, d{\bf x})=
a_{11}{\alpha\over r^3} (\omega^{45}-i\omega^{35})\, .
\ee
The vector potential for a general dipole solution can be written as
follows
\be
({\bf A}, d{\bf x})={\kappa \over 8\pi} {j_{ab}x^b dx^a\over r^3}\,
,
\ee
where
\be
j_{ab}x^b dx^a={ 8\pi \alpha\over \kappa}\left[ a_{10}
\omega^{34}+\sqrt{2} \Re(a_{11})\omega^{45}+\sqrt{2}\Im (a_{11})
\omega^{35}\right]\, ,
\ee
and $a_{10}$, $a_{11}$ are arbirary functions of $u$.

\section{Higher dimensional case}
\n{s6}

Let us discuss first the scalar (electrostatic) equation (\ref{el}) in
the $n-$dimensional Euclidean space $R^n$
\be\n{llap}
\lap \varphi =0\, .
\ee 
To solve this equation it is convenient to decompose the potential
$\varphi$ into the scalar spherical harmonics  \cite{RuOr}
\be\n{yy}
Y^l=r^{-l}{\cal Y}^l\hh
{\cal Y}^l=C_{c_1\ldots c_{l-1}}  x^{c_1}\ldots x^{c_{l-1}}\, ,
\ee
where $C_{c_1\ldots c_{l-1}}$ is a symmetric traceless rank-$l$
tensor. It is easy to see that the number of linearly
independent components of coefficients $C_{c_1\ldots c_{l-1}}$ is
\be
d_0(n,l)={(l+n-3)! (2l+n-2)\over l! (n-2)!}\, . 
\ee
These harmonics are
eigenfunctions of the invariant Laplace operator on a unit sphere
$S^{n-1}$ with eigenvalues $-l(n+l-2)$. For each $l$ there exists
$d_0(n,l)$ linearly independent harmonics. We shall use an index $q$ to
enumerate the independent harmonics. The functions $Y^{lq}$ form a
complete set, so that any smooth function $F$ on $S^{n-1}$ can be
decomposed as
\be
F=\sum_{l=0}^{\infty} \sum_{q} F_{lq}Y^{lq}\, .
\ee 
 
Consider now a special mode $F_{lq}(r)Y^{lq}$. It is a
decreasing-at-infinity solution of (\ref{llap}) if $F_{lq}\sim
r^{-(n+l-2)}$. This can be proved by using the properties of the scalar
spherical harmonics. We demonstrate this directly by using the
relations (\ref{yy}).  

First, it is easy to check that
\be
\lap {\cal Y}^l=0\hh
x^d\pa_d {\cal Y}^l=l {\cal Y}^l\, .
\ee
Using these relations one obtains
\be
\lap (f(r){\cal Y}^l)= (f'' +{(n+2l-1)\over r}f') {\cal Y}^l\, .
\ee
Thus for $f=1/r^{n+2l-2}$ the mode functions $f(r){\cal Y}^l$ obey
the equation (\ref{llap}).  To summarize, a general solution of the
electrostatic equation (\ref{llap}) can be written in the form
\be
\phi=\sum_{l=0}^{\infty} \sum_q {{\cal Y}^{lq}\over r^{n+2l-2}}\, .
\ee
In the gyraton solution (\ref{2.1}) $d_0(n,l)$ independent components of
$C_{c_1\ldots c_{l-1}}$ are arbitrary functions of $u$.

In a similar way, one can obtain solutions of the equations of
magnetostatics in $n-$dimensional Euclidean space by using the  vector
spherical harmonics  \cite{RuOr}. Let us denote
\be
A_a^l=f(r){\cal Y}^l_a\, ,
\ee
\be
{\cal Y}^l_a=C_{abc_1\ldots c_{l-1}} x^b x^{c_1}\ldots x^{c_{l-1}}\, .
\ee
Here $C_{abc_1\ldots c_{l-1}}$ is a $(l+1)$-th-rank constant tensor
which possesses the following properties: it is antisymmetric under
interchange of $a$ and $b$, and it is traceless under contraction of any
pair of indices \cite{RuOr}.

First, let us demonstrate that  $A_a^l$ obeys the gauge
condition
\be\n{g1}
\pa^a A_a^l=0\, .
\ee
Notice that
\be
\pa_a f(r)=f'(r) {x^a\over r}\, .
\ee
Thus
\be
\pa^a A_a^l=f \pa^a {\cal Y}^l_a=0\, .
\ee
The latter equality follows from the fact that when $\pa_a$ is acting
on one of $x$ it effectively produces a contraction of two indices in
$C$ which vanishes. 

In the gauge (\ref{g1}) the magnetostatic field equations (\ref{e1})
reduce to the following equation
\be\n{aa}
\lap A_a^l=0\, .
\ee
It is easy to get
\be
\lap {\cal Y}^l_a=0\hh
x^b \pa_b {\cal Y}_a^l=l {\cal Y}_a^l\, .
\ee
Using these relations one obtains
\be
\lap (f {\cal Y}_a^l)=(f'' +{n+2l-1\over r}f'){\cal Y}_a^l\, .
\ee
Hence ${\cal Y}_a^l$ is a solution of (\ref{aa}) if
\be
f'' +{n+2l-1\over r}f'=0\, .
\ee
Solving this equation we get $f=1/r^{n+2l-2}$. Hence a general
decreasing at infinity solution of the magnetostatic equations in the
$n-$dimensional space (\ref{e1}) can be written as
\be\n{an}
A_a=\sum_{l=1}^{\infty} \sum_{q} {{\cal Y}^{lq}_a\over r^{n+2l-2}}\, .
\ee
Again, we use an index $q$ to enumerate different linearly independent
vector spherical harmonics. The total number of these harmonics for
given $l$ is  \cite{RuOr}
\be
d_1(n,l)={l(n+l-2)(n+2l-2)(n+l-3)!\over (n-3)! (l+1)!}\, .
\ee 
In the gyraton solution (\ref{2.1}) the coefficients $C_{abc_1\ldots
c_{l-1}}$ in the
decomposition (\ref{an}) are arbitrary functions of the retarded time $u$.
For a given solution ${\bf A}$ relation (\ref{psi}) allows one to find
$\psi$.

\section{Summary and discussions}
\n{s7}

The main result of this paper is that the vacuum Einstein equations
for the gyraton-type metric (\ref{2.1}) in an arbitrary number of
spacetime dimensions $D$ can be reduced to linear problems in the
Euclidean $(D-2)-$dimensional space. These problems are: (1) To find
a static electric field $\varphi$ created by a point-like source; (2)
To find a magnetic field ${\bf A}$ created by a point-like source.
The retarded time $u$ plays the role of an external parameter. One
can include $u$-dependence by making the coefficients in the harmonic
decomposition for $\varphi$ and ${\bf A}$ to be arbitrary functions
of $u$. After choosing the solutions of these two problems one can
define $\psi$ by means of equation (\ref{psi}). By substituting
$\Phi=\varphi +\psi$ and ${\bf A}$ into the metric ansatz one obtains
a vacuum solution of the Einstein equations. 

Such a gyraton-like solution has a singularity located at the spatial
point ${\bf x}=0$ during some interval of the retarded time $u$. It
means that the corresponding point-like source is moving with the velocity of
light. Energy $E$ and angular momentum $J_{ab}$ are finite. It was
demonstrated that for given energy and angular momentum the gyraton
can also have other characteristics, describing the deviation of $\Phi$
from spherical symmetry (in the transverse space $R^n$) and the
presence of higher than dipole terms in the multipole expansion of
${\bf A}$. One can interpret such solutions as excitations or
distortions of the gyraton solutions obtained earlier in
\cite{FrFu:05}. 

It should be emphasized that the point-like sources are certainly an
idealization. In \cite{FrFu:05} it was shown that gyraton solutions
can describe the gravitational field of beam-pulse spinning
radiation. In such a description one uses the geometric optics
approximation. For its validity the size of the cross-section of the
beam must be much larger than the wave-length of the radiation. In
the presence of spin $J$ one can expect additional restrictions on
the minimal size of both, the cross-section size and the duration of
the pulse. As usual in physics, one must have in mind that in the
possible physical applications the obtained solution is  valid only
outside some region surrounding the immediate neighborhood of  the
singularity.

The gyraton solutions might be used for study the gravitational
interaction of ultrarelativistic particles with spin. The gyraton
metrics might be also interesting  as possible exact solutions in the
string theory.


\noindent
\section*{Acknowledgments}
\noindent
The authors are grateful to  Alan Coley and Dmitri Fursaev for
stimulating discussions and remarks. This work was supported by the
Natural Sciences and Engineering Research Council of Canada and by
the Killam Trust. The authors also kindly acknowledge the support
from  the NATO Collaborative Linkage Grant (979723).



\end{document}